\documentclass[prl,twocolumn,floatfix,nobibnotes,superscriptaddress,prb]{revtex4}
\usepackage{amsmath}
\usepackage{graphicx}

\begin{document}
\title{Temperature driven $\alpha$ to $\beta$ phase-transformation in Ti, Zr and Hf  from first principles theory combined with lattice dynamics}
\author{Petros Souvatzis}
\affiliation{Department of Physics and Astronomy, Division of Materials Theory Uppsala University,
Box 516, SE-751210, Uppsala, Sweden}
\author{Sergiu Arapan}
\affiliation{Department of Physics and Astronomy, Division of Materials Theory Uppsala University,
Box 516, SE-751210, Uppsala, Sweden}
\affiliation{Institute of Electronic Engineering and Nanotechnologies, Academy of Sciences of Moldova, Academiei 3/3, MD-2028 Chi\c{s}in\u{a}u, Moldova}
\author{Olle Eriksson}
\affiliation{Department of Physics and Astronomy, Division of Materials Theory Uppsala University,
Box 516, SE-751210, Uppsala, Sweden}
\date{\today }
\author{Mikhail I. Katsnelson}
\affiliation{Radboud University Nijmegen, Institute for Molecules
and Materials, NL-6525 AJ Nijmegen, The Netherlands}
\begin{abstract}
Lattice dynamical methods used to predict phase transformations in
crystals typically deal with harmonic phonon spectra and are
therefore not applicable in important situations where one of the
competing crystal structures is unstable in the harmonic
approximation, such as the bcc structure involved in the hcp to
bcc  martensitic phase transformation in  Ti, Zr and Hf. Here we present an
expression for the free energy  that does not suffer from such
shortcomings, and we show by self consistent {\it ab initio}
lattice dynamical calculations (SCAILD),  that the critical
temperature for the hcp to bcc phase transformation in Ti, Zr and Hf,  can be
effectively calculated  from the free energy difference between
the two phases. This opens up the possibility to study
quantitatively, from first principles theory, temperature induced phase
transitions.
\end{abstract}
\pacs{65.40.De, 63.20.Dj, 71.20.Be}

\maketitle

Martensitic phase transformations are common,  both in alloys
frequently used in industry, such as shape memory alloys
\cite{SMA},  and in the the elemental group 3 to 4 transition
metals \cite{WPETRY}, not to mention martensitic transformation
in iron and iron-based alloys,  a crucial phenomenon for
metallurgy \cite{steel}. Thus there exists a substantial interest
both from an industrial, applied and an academic point of view to
develop accurate and effective methods to understand and even
predict martensitic phase-transformations.

The hcp to bcc (or $\alpha$ to $\beta$) transition in Ti, Zr and Hf is a
martensitic phase transformation that has been thoroughly
investigated both from an experimental \cite{WPETRY,RENKEN}  and
theoretical \cite{Henning,Udomsilp, Kristin, Grimvall}
perspective. Recently Hennings {\it et al} developed and used a
classical potential of the modified embedded atom method (MEAM)
\cite{MEAM} to accurately reproduce the phase boundary between the
hcp and bcc structure in Ti. However, there is up to this date a
lack of first principles theoretical studies made of the
martensitic hcp to bcc phase-transformation in Ti, Zr and Hf. The problem is
that anharmonic effects in lattice dynamics~\cite{ktsn} are of
crucial importance for finite-temperature structural phase
transitions, and their quantitative first-principle treatment is a
real challenge.

A straightforward calculation using DFT molecular dynamics
(DFT-MD)\cite{PARINELLO} should in principle be able to reproduce the
bcc to hcp phase transformation in Ti and similar materials, since DFT-MD implicitly include
anharmonic effects. However, DFT-MD is a computationally very
demanding task which makes its use problematic. Instead we will
here exploit the method of self consistent {\it ab initio} lattice
dynamical calculations (SCAILD) \cite{petros1}. Here, we further develop
this method in order to be able to calculate {\it thermodynamic}
properties, such as structural free energy difference (before we
were restricted by the calculations of temperature-dependent
phonon frequencies only
\cite{petros1,petros2,petros3,petros4,petros5}). Since the SCAILD scheme 
is a constrained sampling method, in that it only samples the
lattice dynamical phase-space along the normal mode directions of
commensurate phonons \cite{petros1,petros2}, the SCAILD
calculations are much faster and, thus, much more practical than
the corresponding DFT-MD calculations.

Thus, we propose here a new expression of the free energy of the
SCAILD scheme, and we show that from the atomic configurations and
the phonon density of states produced by the SCAILD calculations,
an accurate measure of the free energy for the different phases
can be obtained.

In order to properly describe temperature driven phase
transformations in general, one must include the interaction
between phonons~\cite{ktsn}. As a result, phonon frequencies turn
out to be temperature dependent which we explore numerically in
this study by means of the SCAILD method
\cite{petros1,petros2,petros3,petros4,petros5}.

The SCAILD method is based on the calculation of Hellman-Feynman
forces of atoms in a supercell. The  method can be viewed as an
extension of the frozen phonon method \cite{FP1}, in which all
phonons with wave vectors $\mathbf{q}$ commensurate with the
supercell are excited together in the same cell by displacing
atoms situated  at the undistorted positions
$\mathbf{R}+\mathbf{b}_{\sigma}$, according to
 $\mathbf{R}+\mathbf{b}_{\sigma}  \rightarrow \mathbf{R}+\mathbf{b}_{\sigma} + \mathbf{U}_{\mathbf{R}\sigma}$, where the displacements are given by
\begin{equation}
\mathbf{U}_{\mathbf{R}\sigma}
= \frac{1}{\sqrt{N}}\sum_{\mathbf{q},s}\mathcal{A}_{\mathbf{q}s}^{\sigma}
\mathbf{\epsilon}_{\mathbf{q}s}^{\sigma}e^{i\mathbf{q}(\mathbf{R}+\mathbf{b}_{\sigma})}.\label{eq:SUPERPOS}
\end{equation}
Here $\mathbf{R}$ represent the $N$ Bravais lattice sites of the
supercell, $\mathbf{b}_{\sigma}$ the position of atom $\sigma$
relative to this site, $\mathbf{\epsilon}_{\mathbf{q}s}^{\sigma}$
are the phonon eigenvectors corresponding to the phonon mode, $s$,
and the mode amplitude $\mathcal{A}_{\mathbf{q}s}^{\sigma}$  is calculated from the
different phonon frequencies
 $\omega_{\mathbf{q}s}$ through
\begin{equation}
 \mathcal{A}_{\mathbf{q}s}^{\sigma} =\pm \sqrt{
\frac{\hbar}{M_{\sigma}\omega_{\mathbf{q}s}} \Big
(\frac{1}{2}+n_{\mathbf{q}s}\Big )},
\label{eq:AMPL}
\end{equation}
where $n_{\mathbf{q}s}=n(\frac{\omega_{\mathbf{q}s}}{k_{B}T})$, with $n(x)=1/(e^{x}-1)$, are the phonon 
occupational numbers, $M_{\sigma}$ the atomic masses and $T$ is the temperature of the system. The phonon frequencies, $\omega_{\mathbf{q}s}$, are defined through the  variational derivative of the total energy with
respect to the occupation numbers
\begin{eqnarray}\label{eq:FOURIER}
\hbar \omega _{\mathbf{q}s} =\frac{\delta E_{tot}}{\delta n_{\mathbf{q}s} }= \qquad \qquad \nonumber \\  
\sum_{\sigma}\frac{\delta E_{tot}}{\delta \mathcal{A}_{\mathbf{q}s}^{\sigma}} 
\frac{\delta \mathcal{A}_{\mathbf{q}s}^{\sigma}}{\delta n_{\mathbf{q}s}}= 
 -\frac{\hbar}{\omega_{\mathbf{q}s}}\sum_{\sigma}\frac{\mathbf{\epsilon}_{\mathbf{q}s}^{\sigma} \cdot \mathbf{F}_{\mathbf{q}}^{\sigma}}{\mathcal{A}_{\mathbf{q}s}^{\sigma}M_{\sigma}},
\end{eqnarray}
obtained through  the Fourier transform
$\mathbf{F}_{\mathbf{q}}^{\sigma}$ of the forces acting on the
atoms in the supercell.

Due to the simultaneous presence of all the commensurate phonons
in the same force calculation, the interaction between different
lattice vibrations are  taken into account and the phonon
frequencies given by  Eq. (\ref{eq:FOURIER}) are thus renormalized
by the very same interaction.

By alternating between calculating the forces on the displaced
atoms and calculating new phonon frequencies and new displacements
through Eqs. (\ref{eq:SUPERPOS})-(\ref{eq:FOURIER}) the phonon
frequencies are calculated in a self consistent manner. For more
details on the SCAILD method we refer to Refs.
\onlinecite{petros1,petros2,petros3}. It should be mentioned that
we do not consider here the phonon decay processes (see, e.g.,
Ref.~\onlinecite{CaSr} and references therein).

The free energy as a function of volume, V, and temperature for
the bcc and hcp structures can be calculated  through the
expression
 \begin{equation}
 F(T,V) = U_{0}(V)+F_{ph}(V,T)+F_{el}(V,T),
 \label{eq:F1}
 \end{equation}
where $U_{0}$ is the static ground state energy of the respective
structures at $T = 0$ K (i.e  without any phonons excited and
temperature excitations of electronic states ),  $F_{ph}$ is the
free energy of the phonons and $F_{el}$ is the free energy of the
electrons. The temperature dependent parts of the free energy can
be found as
\begin{eqnarray}
 F_{ph}(V,T) + F_{el}(V,T)=   \qquad \qquad  \qquad \qquad \qquad \nonumber \\
 \frac{1}{N_{I}}\sum_{\{\mathbf{U}_{\mathbf{R}}\}}\Delta F^{*}(\{\mathbf{U}_{\mathbf{R}}\},V,T)
 +\frac{3}{2}k_{B}T-TS_{ph}(V,T).
 \label{eq:F2}
\end{eqnarray}
Here $\Delta F^{*}$ is the change in free energy relative to the
ground state energy $U_{0}$, caused by the phonon induced atomic
displacements described by Eq. (\ref{eq:SUPERPOS}), and thermal
excitations of the electronic states.  The sum on the right-hand
part of Eq. (\ref{eq:F2}) is over the different atomic
configurations, $\{\mathbf{U}_{\mathbf{R}}\}$,  generated
throughout  the SCAILD self consistent run. Since  $\Delta F^{*}$
are calculated at atomic configurations accommodating the
different  frozen phonon superposition of Eq. (\ref{eq:SUPERPOS}),
$\Delta F^{*}$ not only contains the finite temperature
contribution to the electronic free energy for a given atomic
configuration, but also, the potential energy provided by the
frozen lattice waves, i.e the potential energy of the phonons at a
particular phonon superposition  \cite{Arapan}. The phonon kinetic energy is
given by $3k_{B}T/2$ per atom which means that atomic motion is
considered as classical; typically, temperatures of structural
phase transformations are higher than the Debye temperature, thus,
this approximation is well justified.

In practice, the sum of the  finite temperature electron free
energy and phonon potential energy, $\Delta F^{*}$, was obtained
by calculating the total free energy of the corresponding atomic
configuration using a Fermi-Dirac temperature smearing of the
Kohn-Sham occupational numbers \cite{VASP}, and then subtracting
the static potential energy $U_{0}$ (Here $U_{0}$ is  calculated
with the tetrahedron method to provide a good reference to the
temperature excited electronic states). The number of
configurations, $N_{I}$, used for each volume and temperature was
typically 400.

Another problem is how to calculate the phonon entropy. We will assume
that it depends on the phonon occupation numbers $n_{\mathbf{q}s}$ in
the same way as for noninteracting bosons:
\begin{equation}
S_{ph}=k_B\sum\limits_{\mathbf{q}s} \left[ \left( 1+n_{\mathbf{q}s} \right)
\ln \left( 1+n_{\mathbf{q}s} \right) -n_{\mathbf{q}s} \ln n_{\mathbf{q}s} \right]
\label{eq:entropy}
\end{equation}
 This is the only entropy expression
consistent with Eq. (\ref{eq:AMPL}) and (\ref{eq:FOURIER}),
which  can be proved in the exact same manner as was done for  fermions in the 
Landau theory of a normal Fermi liquid \cite{landau}. Thus the SCAILD scheme, together with a free energy defined through  Eqs. (\ref{eq:F2}) and
(\ref{eq:entropy}), constitutes nothing but a theory of a ``normal Bose
liquid''. Expression (\ref{eq:entropy}) can be written in terms of the
phonon density of states, $g(\omega)$,  produced by a converged
SCAILD calculation, and is given by
 \begin{eqnarray}
 TS_{ph}(V,T) = \qquad \qquad  \qquad \qquad \qquad  \nonumber \qquad \qquad  \qquad \quad \\
 \int_{0}^{\infty}d\omega g(\omega,V,T)\hbar\omega\Big [n(\frac{\hbar\omega}{k_{B}T})-\frac{k_{B}T}{\hbar\omega}\ln \Big (1-e^{-\frac{\hbar\omega}{k_{B}T}}\Big ) \Big ].
 \label{eq:TS}
  \end{eqnarray}
Here the phonon frequencies used to calculate the phonon density of states, $g(\omega)$, are the normal mode configurational mean values
\begin{equation}
\langle \hbar \omega _{\mathbf{q}s} \rangle=\frac{1}{N_{I}}\sum_{\{\mathbf{U}_{\mathbf{R}}\}}\frac{\delta E_{tot}(\{\mathbf{U}_{\mathbf{R}}\},V,T)}{\delta n_{\mathbf{q}s} }.
 \end{equation}

It should be stressed that the partitioning of the free energy
through Eqs. (\ref{eq:F1}), (\ref{eq:F2}) and (\ref{eq:TS}) has
been chosen to maximize both the accuracy of the phonon potential
energy, which in the form of Eqn.(\ref{eq:F2}) take into account
anharmonicity up to infinite order, and the phonon entropy, which
in the form given by Eqn.(\ref{eq:TS}) is accurate to leading order in
anharmonic perturbation theory \cite{cochran}.

\begin{figure}[tbp]
\begin{center}
\includegraphics*[angle=0,scale=0.30]{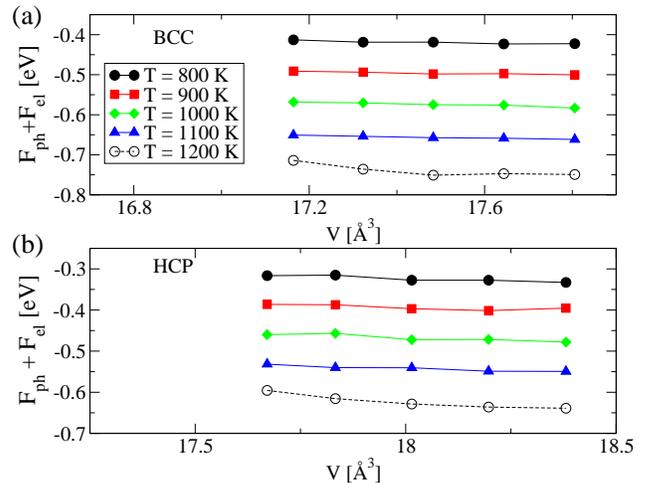}
\caption{(Color online) The calculated free energy at different volumes and temperatures for Ti in the bcc phase (a),  and the  hcp
phase (b).}
\label{fig:Free1}
\end{center}
\end{figure}

The phonon density of states and the corresponding  free energies
for the hcp and bcc structures were calculated at up to five different
temperatures, and at each temperature, SCAILD calculations were
performed at up to five different volumes. As an example of the typical data obtained from the calculations, the resulting Titanium free
energies obtained through Eq. (\ref{eq:F2}), are shown in Fig.
\ref{fig:Free1}.

As regards the other computational details of the force
calculation we used the VASP package \cite{VASP}, within the
generalized gradient approximation (GGA). The projector-augmented wave (PAW) potentials
required energy cutoffs of 232 eV. The Ti(4s,3d), Zr(4s,4p,5s,4d) and Hf(6s,5d) levels were treated as valence electrons. The k-point mesh was a
$5\times5\times5$ Monkhorst-Pack grid in the bcc phase
calculations. In the hcp phase calculations $6\times6\times6$
gamma centered mesh was used.  In order to include the electron
entropy in the calculations, Fermi-Dirac temperature smearing were
applied to the Kohn-Sham occupational numbers. The bcc and hcp
supercells used were obtained by increasing the bcc primitive cell
4 times  and  the hcp primitive cell 3 times, along  the
respective bcc and hcp primitive lattice vectors. 
The sizes of the supercells where chosen such that to ensure a sufficient decay of the interatomic force constant within the supercell,
permitting a proper sampling of the lattice dynamical phase space \cite{petrosAu}.
For the calculations of the static potential
energy, $U_{0}$ in Eqn.(\ref{eq:F1}), the all-electron full-potential linearized augmented-plane wave (FP-LAPW) package ELK \cite{elk} was used within the GGA approximation. 
This was found necessary to ensure a high accuracy of the zero temperature part of the bcc - hcp energy difference.
An energy cutoff of 270 eV  together with a
$24\times24\times24$  k-point mesh and a
$24\times24\times15$  k-point mesh were used for the bcc
and hcp structures, respectively, in these calculations the Methfessel-Paxton integration scheme was 
used with a 0.2 eV smearing  of the Kohn-Sham eigenvalues.

\begin{figure}[tbp]
\begin{center}
\includegraphics*[angle=0,scale=0.31]{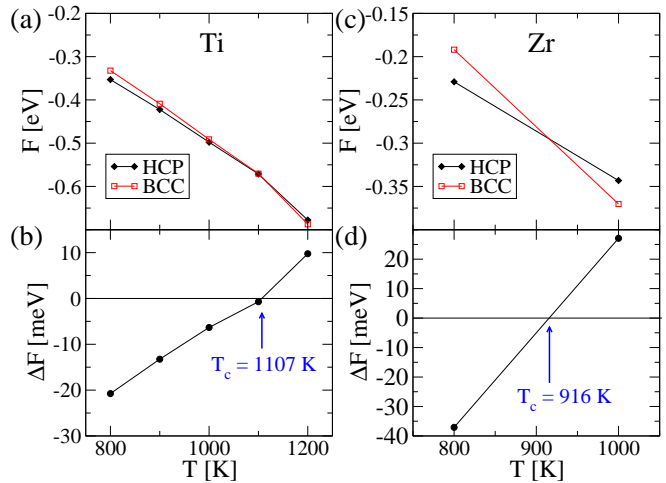}
\caption{(Color online) In (a) and (c) the calculated free energy of the bcc and hcp phase, in Ti and Zr, respectively. 
In (b) and (d)  the calculated free energy difference, $\Delta F= F_{hcp}-F_{bcc}$, in Ti and Hf, respectively}
\label{fig:Free2}
\end{center}
\end{figure}
\begin{figure}[tbp]
\begin{center}
\includegraphics*[angle=0,scale=0.3]{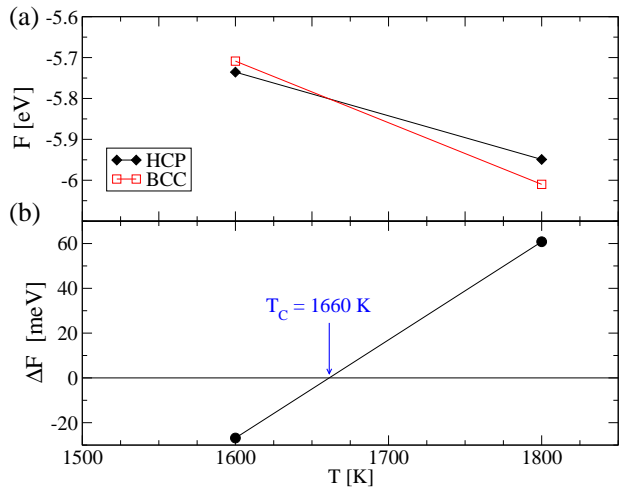}
\caption{(Color online) In (a) the calculated free energy in Hf for the bcc and hcp phase represented by red
empty squares  and  filled black diamonds, respectively.
In (b)  the calculated free energy difference in Hf, $\Delta F= F_{hcp}-F_{bcc}$.}
\label{fig:disp}
\end{center}
\end{figure}

For each temperature the free energy obtained through Eq.
({\ref{eq:F2}) was fitted to a first order polynomial in $V$ (for typical data see Fig. \ref{fig:Free1}). Then by using these first order
fits together with  Eq. (\ref{eq:F1}), the total free energy at
each temperature was obtained through  minimization with respect
to volume. In Fig. \ref{fig:Free2}(a), (c) and \ref{fig:disp} (a), the minimized  free energy at
each temperature is displayed for the bcc and hcp structure in Ti, Zr and Hf, respectively.
In Fig. \ref{fig:Free2}(b), (d) and \ref{fig:disp} (b) the free energy difference between the
structures is displayed for Ti, Zr and Hf, respectively. 

The temperature driven hcp to bcc
phase-transformation in Ti can bee seen to occur at $T\sim 1100$ K, which
is reasonably  close to the experimentally observed phase
transition temperature of 1155 K \cite{expbcc}, whereas in Zr and Hf, the transition is predicted  to occur at T$\sim$ 920 K and T$\sim$1660 K, respectively. The theoretical 
estimates of  the transition temperature in Zr and Hf, with their respective experimental data of 1135 K and 2015 \cite{WPETRY}, does not correspond to experimental data 
as well as in the case of Ti, however, given that first principles calculations (at $T=0$), with current exchange
and correlations functionals, have a problem in resolving energy
differences between different crystallographic phases better than
$\sim$10 meV/atom, one may not expect from any first
principles based theory, like the SCAILD method used here, to
reproduce temperature induced phase transitions with an accuracy
better than  a few hundred Kelvin. 

 It should be noted here that the presently used free energy expression differs significantly from the previously used
expression from quasi-harmonic theory (e.g. used in Ref.\cite{petros2} ) in that it takes into account the anharmonic contributions  to the lattice dynamical potential energy without 
projecting it down on a quasi-harmonic description as in Ref.\cite{petros2}.  In fact, using the old free energy expression together with the 
first principles inter atomic forces calculated in this work, results in the hcp phase in Ti being $\gtrsim$ 90 meV
lower in free energy compared to the free energy of  the bcc phase throughout   the entire temperature intervall $800K<T<1200K$. 


In summary, we introduce here an expression for the free energy which is
applicable also for highly anharmonic crystals. The free energy
expression can readily be interfaced with lattice dynamic methods,
for instance the SCAILD technique, and we show that the
temperature induced hcp $\rightarrow$ bcc transition of Ti, Zr and Hf  can be
reproduced by theory. Theory puts the transition temperature in Ti, Zr and Hf  to
1100 K, 920 K and 1660 K, respectively. 

We would like to thank the Swedish National Infrastructure for
Computing (SNIC) for the allocation of computational time at NSC, HPC2N and C3SE that
made this work possible.  Support from VR, ERC, the KAW foundation and ESSENCE is acknowledged.


\end{document}